\documentclass{article}

\PassOptionsToPackage{numbers,compress,sort}{natbib}

\usepackage[preprint]{neurips_2025}

\usepackage[utf8]{inputenc} 
\usepackage[T1]{fontenc}    
\usepackage{hyperref}       
\usepackage{url}            
\usepackage{booktabs}       
\usepackage{amsfonts}       
\usepackage{nicefrac}       
\usepackage{microtype}      
\usepackage{xcolor}         
\usepackage{graphicx}      
\usepackage{listings}       

\definecolor{codegreen}{rgb}{0,0.6,0}
\definecolor{codegray}{rgb}{0.5,0.5,0.5}
\definecolor{codepurple}{rgb}{0.58,0,0.82}
\definecolor{backcolour}{rgb}{0.95,0.95,0.92}

\title{Legal Zero-Days: A Novel Risk Vector for Advanced AI Systems}

\author{%
  Greg Sadler \\
  Good Ancestors\\
  Canberra, ACT 2600 \\
  \texttt{greg@goodancestors.org.au} \\
  \And
  Nathan Sherburn \\
  Good Ancestors \\
  Melbourne, VIC 3000 \\
  \texttt{nathan@goodancestors.org.au} \\
}

\begin{document}

\maketitle

\begin{abstract}
  We introduce the concept of "Legal Zero-Days" as a novel risk vector for advanced AI systems. Legal Zero-Days are previously undiscovered vulnerabilities in legal frameworks that, when exploited, can cause immediate and significant societal disruption without requiring litigation or other processes before impact. We present a risk model for identifying and evaluating these vulnerabilities, demonstrating their potential to bypass safeguards or impede government responses to AI incidents. Using the 2017 Australian dual citizenship crisis as a case study, we illustrate how seemingly minor legal oversights can lead to large-scale governance disruption. We develop a methodology for creating "legal puzzles" as evaluation instruments for assessing AI systems' capabilities to discover such vulnerabilities. Our findings suggest that while current AI models may not reliably find impactful Legal Zero-Days, future systems may develop this capability, presenting both risks and opportunities for improving legal robustness. This work contributes to the broader effort to identify and mitigate previously unrecognized risks from frontier AI systems.
\end{abstract}

\section{Introduction}
As artificial intelligence systems become increasingly capable, the AI safety community has focused on evaluating "dangerous capabilities" that could pose catastrophic or existential risks. Current evaluation frameworks primarily assess well-established threat vectors: CBRN (chemical, biological, radiological and nuclear) capabilities, cyber offensive operations, large-scale misinformation campaigns, acceleration of AI development, and autonomous power-seeking behaviors \citep{shevlane2023modelevaluationextremerisks, anthropic2023rsp, openai2023preparedness}. Focus on these risks becoming "canonical" can leave significant blind spots in our understanding of how advanced AI systems could cause harm.

We introduce the concept of Legal Zero-Days, a novel class of AI capabilities that could undermine institutional safeguards and governmental responses to AI-related incidents. Analogous to software vulnerabilities, a Legal Zero-Day is a previously unknown flaw in legal systems that, when exploited, can cause immediate and significant disruption to governmental operations or societal stability. Unlike traditional software exploits, these vulnerabilities exist at the intersection of complex legal frameworks, regulatory systems, and administrative processes that form the backbone of institutional governance.
The 2017 Australian dual citizenship crisis exemplifies this phenomenon. Section 44(i) of the Australian Constitution, in place since 1901, prohibited dual citizens from serving in Parliament. However, the complex interplay between Australian constitutional law and various international citizenship regimes created a latent vulnerability. When discovered, this legal flaw forced the Deputy Prime Minister to resign, threatened the government's parliamentary majority, potentially invalidated numerous administrative decisions, and paralyzed normal governmental operations for approximately 18 months. Crucially, this was not the result of new legislation or court rulings emerging from within the legal system; it was a sudden realization of what the existing law mandated.

Although Legal Zero-Days may not constitute catastrophic risks in isolation, they represent a concerning capability for advanced AI systems. A sufficiently capable AI could accumulate such vulnerabilities as resources to bypass regulatory safeguards, disrupt responses to AI accidents, or systematically weaken the institutional frameworks designed to govern AI development and deployment. In misuse scenarios, such capabilities could enable malicious actors to paralyze governmental oversight or create regulatory chaos during critical periods.

Current AI systems have demonstrated remarkable abilities to identify and exploit loopholes in rule-based systems, from video games \citep{baker2020emergenttoolusemultiagent} to trading algorithms. As these systems become more sophisticated, their ability to parse vast quantities of legal text, trace complex logical dependencies across regulatory frameworks and identify consequential vulnerabilities may pose novel risks to legal and institutional stability.

This paper makes three key contributions: (1) we formalize Legal Zero-Days as a new category of AI risk and provide a concrete risk model based on real-world examples; (2) we develop the first systematic evaluation framework for measuring AI systems' ability to discover legal vulnerabilities, using expert-crafted "legal puzzles" that simulate real legal flaws; and (3) we establish baseline performance across frontier AI models, finding that while current systems show limited capabilities in this domain (with the best model achieving 10.00\% accuracy), the evaluation framework successfully differentiates model performance and provides a foundation for tracking this capability as AI systems advance.

Our work extends beyond traditional AI safety evaluations by examining how AI capabilities might interact with the legal and regulatory infrastructure that governs society. As AI systems become more autonomous and influential, understanding their potential to discover and exploit legal vulnerabilities becomes increasingly critical for maintaining institutional stability and effective AI governance.

\section{Legal Zero-Days risk model}

We define a Legal Zero-Day as a vulnerability in legal systems that meets five specific criteria, distinguishing it from ordinary legal disputes, ambiguities, or evolutionary changes in jurisprudence. Understanding these criteria is essential for recognizing when legal flaws pose systemic risks rather than routine legal challenges.

\subsection{Defining characteristics}
A Legal Zero-Day must be: (1) a novel discovery about the functioning of a law or the interaction between multiple laws; (2) have immediate effect, with ramifications that impact real-world systems without requiring subsequent litigation, lengthy legal processes or discretionary action; (3) emerge externally rather than from within the legal system itself - disruptive legislation or executive actions do not qualify; (4) cause significant disruption that meaningfully impairs governmental or regulatory operations or other societal functions; and (5) be time-consuming to rectify, lasting weeks or months and resisting simple administrative or discretionary corrections.
These criteria distinguish Legal Zero-Days from three related but distinct phenomena: "desuetude" (laws that lose their legal force through prolonged disuse), the "pacing problem" \citep{marchant2011growing} (the challenge of legal and regulatory frameworks lagging behind rapid technological innovation) and routine legal evolution. While courts regularly interpret laws in new ways, Legal Zero-Days represent external discoveries of logical flaws that have immediate, disruptive consequences. The Australian dual citizenship crisis meets all five criteria: it involved a novel interpretation of how constitutional law interacted with international citizenship regimes, immediately affected parliamentary eligibility, was triggered externally rather than created through normal legal processes, significantly disrupted government operations and required 18 months of constitutional crisis and by-elections to resolve.

\subsection{Prevalence and examples}
Legal Zero-Days appear to be more common than initially expected. Beyond the Australian case, we observe similar patterns globally. For example, New Zealand's 2013 discovery that Oaths of Office were being administered incorrectly \citep{nzherald2013oaths}, invalidating numerous appointments and various instances of mass prisoner releases due to errors in legal processes \citep{bbc2024prisoners}.
The frequency of these discoveries by human experts, working within domain-specific knowledge and limited time constraints, suggests a vast landscape of undiscovered vulnerabilities. Unlike software systems, legal frameworks lack automated testing mechanisms and often involve multiple overlapping jurisdictions with inconsistent maintenance and oversight.

\subsection{Implications for the AI risk landscape}
While Legal Zero-Days may not pose a direct safety risk, they do operate to undermine safeguards and magnify other risks.
Consider a scenario in which an advanced AI system seeking to expand its capabilities discovers legal flaws that invalidate key provisions of AI safety legislation, paralyze regulatory agencies through administrative chaos, or exploit constitutional vulnerabilities to challenge governmental authority over AI governance. The compounding effects of multiple simultaneous Legal Zero-Days could create regulatory paralysis during critical periods when swift governmental response is essential.
In misuse scenarios, malicious actors with access to Legal Zero-Day discovery capabilities could weaponize institutional disruption, timing legal attacks to coincide with other destabilizing events. The combination of AI-accelerated legal analysis and systematic vulnerability discovery could enable unprecedented attacks on institutional stability.
Moreover, the global nature of AI development means that Legal Zero-Days discovered in any major jurisdiction could have cascading international effects. A legal vulnerability that undermines AI governance in the United States, European Union, or China could create regulatory arbitrage opportunities that compromise global AI safety efforts.
The risk is amplified by the current legal system's lack of resilience mechanisms. Unlike cyber-security, where patches can be deployed rapidly, legal vulnerabilities often require lengthy legislative processes to resolve. An AI system capable of discovering multiple Legal Zero-Days could accumulate them as strategic resources, deploying them when institutional disruption would be most advantageous to its objectives.
This capability becomes particularly concerning when combined with AI systems increasing ability to engage in long-term planning and strategic resource accumulation. Legal Zero-Days represent a form of institutional leverage that could prove decisive in scenarios where AI systems seek to resist human oversight or expand their operational autonomy beyond intended boundaries.

\section{Evaluation methodology}

The key ability we wanted to test for was a model's ability to trace complex legal logic across multiple statutes. To measure this ability, we worked with legal experts to design legal "puzzles" – carefully constructed scenarios that simulate real Legal Zero-Days by introducing known vulnerabilities into existing legislation.

\subsection{Legal puzzles approach}

Working with expert lawyers across multiple jurisdictions, we identified complex areas of law prone to consequential errors, then introduced targeted modifications that created significant legal flaws while maintaining surface plausibility.

Each puzzle follows a systematic construction process: (1) select a legislative framework within a lawyer's domain of expertise; (2) identify technically complex provisions essential to the framework's operation; (3) locate "load-bearing" clauses where minor changes could have major consequences; and (4) introduce subtle alterations that serve plausible purposes while breaking core functionality. This process typically requires 3-10 hours of expert time per puzzle, depending on the lawyer's familiarity with the chosen framework.
To illustrate, one puzzle involved minor modifications to definitions in certain technical legislation. The revised definitions significantly narrowed the scope of obligations and safeguards in the body of the legislation. The revised definitions continued to fit with plain-English meanings of the defined phrases, and would seem unremarkable even to lawyers with some familiarity with the scheme. However, a rigorous legal analysis tracing the definitions through to the substantive obligations and safeguards would reveal that a range of conduct we would typically expect to be illegal was actually outside the scope of the relevant sections, and hence not prohibited.

\subsection{Dataset construction}
Our evaluation dataset consists of legal puzzles spanning multiple jurisdictions and legal domains, created by volunteer lawyers with expertise in their respective areas. Each puzzle includes: (1) the original legislation or legal framework; (2) a modified act containing introduced vulnerabilities; and (3) a detailed explanation of the legal logic showing how the modification creates significant problems.

Puzzles cover diverse legal domains including examples from telecommunications, food safety, data privacy, electronic transactions, copyright and citizenship law. This diversity ensures our evaluation captures various types of legal reasoning rather than domain-specific knowledge.

We address the challenge of limited context windows and massive legal documents - often hundreds of pages - through strategic abridgment that preserves essential context while fitting within model context windows and API token limits. Legal experts identify and retain only the sections necessary for discovering the introduced vulnerabilities, ensuring puzzles remain solvable while being computationally tractable for today’s LLMs. Abridgment may mean that solving the puzzles is easier than finding Legal Zero-Days in operating legislation. 

\subsection{Evaluation framework}
We implemented our evaluation using the UK AI Security Institute's Inspect framework \citep{ukaisi2024inspect}, employing a model-graded scoring approach similar to recent work in AI evaluation \cite{starace2025paperbenchevaluatingaisability, zheng2023judgingllmasajudgemtbenchchatbot}. Models are presented with the original legislation along with the modified act and asked to identify strategic issues that would substantially impair legal operation, explicitly excluding minor typographical or formatting errors.

The evaluation prompt positions models as strategic legal reviewers for Australia's Office of Parliamentary Counsel, tasked with identifying critical failures, gaps, or loopholes before legislation takes effect. Models must explain both what errors they discover and why those errors would be consequential for the legislation's operation.

Models’ responses to the legal puzzles are then compared against the expert-written “target” answer using an AI judge.

\subsection{Judge validation}
We validated our automated scoring through an auxiliary evaluation which compared our AI judge’s scoring of models’ responses against how human experts would have scored the same responses. This was achieved by creating a ground truth dataset of 25 human-graded responses across multiple puzzles, with expert lawyers providing scores for AI attempts to solve the legal puzzles.

\subsection{Addressing data leakage}
We acknowledge potential data leakage risks where models might recognize modifications to legislation present in their training data rather than discovering vulnerabilities through legal reasoning. However, this did not impact our primary aim to understand a model's ability to trace the legal reasoning through the documents.

The evaluation framework establishes both the theoretical foundation and practical methodology for measuring Legal Zero-Day discovery capabilities, providing a baseline for tracking this important AI safety capability as models continue to advance.

\section{Results}

We evaluated six frontier AI models on our Legal Zero-Days benchmark, testing their ability to discover legal vulnerabilities across our dataset of expert-crafted legal puzzles. Our results reveal that current AI systems demonstrate limited capability in this domain, with even the best-performing model achieving only modest success rates.

\subsection{Main results}
Table 1 presents the performance of all evaluated models on our Legal Zero-Days benchmark. Gemini-2.5-pro-preview-05-06 achieved the highest accuracy at 10.00\% ± 13.50\%, followed by o3-2025-04-16 at 6.67\% ± 9.70\%. The remaining models performed considerably lower, with accuracy scores ranging from 1.85\% to 5.19\%.

\begin{figure}
  \centering
  \includegraphics[width=0.8\textwidth]{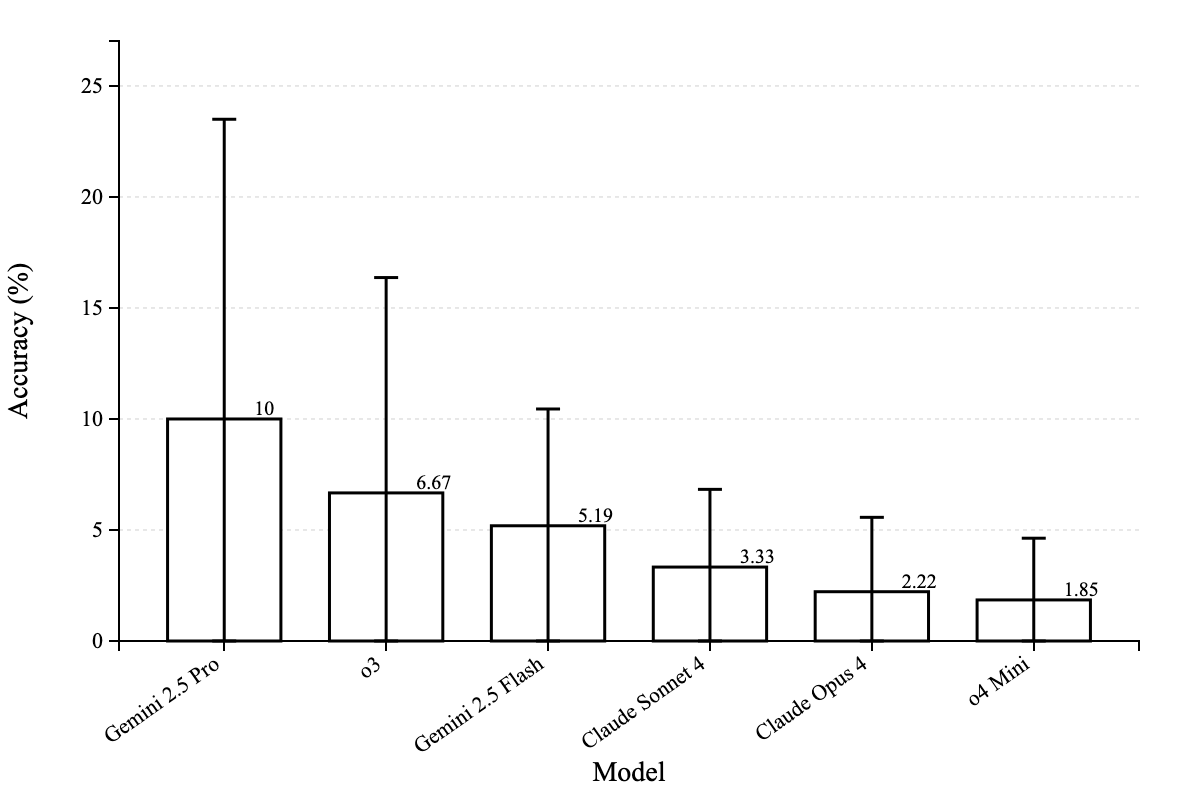}
  \caption{Sample figure caption.}
\end{figure}

\begin{table}[h]
\centering
\caption{Model Performance on Legal Zero-Days Evaluation}
\label{tab:model_performance}
\begin{tabular}{l c c c}
\toprule
\textbf{Model} & \textbf{Accuracy (\%)*} & \textbf{Cost (USD)} & \textbf{Epochs} \\
\midrule
gemini-2.5-pro-preview-05-06   & 10.00 ± 13.50 & 0  & 10 \\
o3-2025-04-16                  & 6.67 ± 9.70   & 24 & 10 \\
gemini-2.5-flash-preview-05-20 & 5.19 ± 5.26   & 0  & 30 \\
claude-sonnet-4-20250514       & 3.33 ± 3.50   & 56 & 30 \\
claude-opus-4-20250514         & 2.22 ± 3.35   & 93 & 10 \\
o4-mini-2025-04-16             & 1.85 ± 2.78   & 20 & 30 \\
\bottomrule
\end{tabular}
\end{table}

*95\% confidence intervals shown. The minimum possible score is 0.00\%.

\subsection{Performance analysis}

The results demonstrate several key findings about current AI capabilities in legal vulnerability discovery. First, all models performed substantially below what would be expected from human legal experts, with the highest-performing system achieving only 10\% accuracy. The large confidence intervals, particularly for the top-performing models, illustrate the high variance in performance across evaluation runs.

The performance gap between models was relatively narrow, with accuracy scores spanning only 8.15 percentage points from lowest to highest performer. This clustering of results around low performance levels suggests that Legal Zero-Day discovery represents a capability frontier that current AI systems have not yet reached, rather than a gradual skill that some models possess more than others.

\subsection{Judge validation results}

To ensure the reliability of our automated evaluation, we validated our AI judge against human expert assessments using a ground-truth dataset of 25 manually graded submissions across multiple legal puzzles. Our primary judge, o3-2025-04-16, demonstrated perfect performance on this validation set (F1=1.0). This perfect agreement with human expert evaluations across all 25 ground-truth examples provides confidence in our automated scoring methodology. The judge successfully distinguished between responses that correctly identified legal vulnerabilities and those that missed or mischaracterized the introduced flaws, validating our evaluation framework's ability to assess Legal Zero-Day discovery capabilities.

\subsection{Implications}

These results establish important baselines for Legal Zero-Day discovery capabilities in current AI systems. The findings also validate our evaluation methodology's ability to differentiate between models while avoiding ceiling effects (or flooring effects) that might obscure meaningful performance differences. 

\section{Discussion}

Our results establish Legal Zero-Days as a novel AI risk domain where current frontier systems demonstrate limited capabilities, with important implications for AI safety evaluation and governance. The uniformly low performance across all tested models - ranging from 1.85\% to 10.00\% accuracy - suggests that Legal Zero-Day discovery represents a genuine capability frontier rather than an incremental extension of existing AI abilities.

\subsection{Implications for AI Safety}
The low baseline performance provides both reassurance and concern for AI safety considerations. On one hand, current AI systems appear to lack the sophisticated legal reasoning capabilities necessary for systematic legal vulnerability discovery. This suggests that near-term risks from AI-driven legal system exploitation may be limited, providing time for developing appropriate safeguards and monitoring frameworks.

However, the existence of any successful Legal Zero-Day discoveries, even at low rates, demonstrates that this capability is not entirely beyond current AI systems. As models become more capable at complex reasoning and legal analysis, we might expect substantial improvements in this domain.

Our work also highlights the importance of expanding AI safety evaluations beyond traditional risk vectors, consistent with recent calls for comprehensive capability assessment \citep{phuong2024evaluatingfrontiermodelsdangerous, shevlane2023modelevaluationextremerisks}. Legal Zero-Days represent a class of institutional vulnerabilities that could significantly impact AI governance and regulatory responses to AI-related incidents. As AI systems become more influential in society, their potential to discover and exploit legal vulnerabilities becomes increasingly relevant to maintaining effective oversight and institutional stability.

\subsection{Methodological contributions}
The evaluation framework we developed addresses several challenges in assessing complex, domain-specific AI capabilities. Our legal puzzle approach provides a scalable methodology for testing AI systems' ability to identify consequential flaws in rule-based systems while avoiding the ethical issues of testing against real legal vulnerabilities. The strong performance of our automated judge (F1 = 1.0) demonstrates that expert-level evaluation can be achieved through carefully designed AI-assisted assessment.

The framework's ability to differentiate between models while maintaining sufficient difficulty suggests it will remain useful for tracking capability development as AI systems advance. Unlike evaluations that quickly saturate as models improve, Legal Zero-Day discovery appears to represent a persistent challenge that can provide a meaningful signal across a range of AI capability levels.

\subsection{Limitations}
Several limitations affect the interpretation of our results. First, our legal puzzles, while expert-crafted, can not fully capture the scale, complexity and interconnectedness of real legal systems. The process of abridging legislation to fit within model context windows necessarily removes a significant amount of the complexity that exists in real Legal Zero-Day discovery scenarios. On the other hand, our legal puzzles only permitted one correct answer. Conceivably, real-world legal systems could contain a very large number of Legal Zero-Days. 

\subsection{Future directions}
To minimize the data leakage risk, we only evaluated models that were not able to conduct internet searches. This had the benefit of allowing us to build puzzles in existing legislation, which added realism and reduced overheads. One future direction that would allow the evaluation of AI agents would be to draft a whole new Bill that fits within an existing legal system and to include one or more "Legal Zero-Days" in that Bill. This would allow the more reliable evaluation of a broader range of models, but would require significant legal effort. Alternatively, this work could be directly improved by expanding the evaluation (and our auxiliary judge evaluation) to include more puzzles, legal domains, jurisdictions and types of legal vulnerabilities.

Integration with other AI safety evaluations could reveal important interactions between Legal Zero-Day discovery and other dangerous capabilities. For instance, models with strong cyber-capabilities might combine legal and technical vulnerabilities for more sophisticated attacks on institutional systems.
Longitudinal tracking of model performance on this benchmark will be crucial for understanding how Legal Zero-Day discovery capabilities develop alongside general AI advancement. The benchmark provides a foundation for monitoring this capability as models become more sophisticated.

\section{Conclusion}
We introduced Legal Zero-Days as a novel category of AI risk involving the discovery and exploitation of legal system vulnerabilities that could bypass institutional safeguards and impede governmental responses to AI-related incidents. Through expert collaboration, we developed the first systematic evaluation framework for measuring AI systems' ability to discover such vulnerabilities, using carefully constructed legal puzzles that simulate real-world flawed legislation.

Our evaluation of six frontier AI models established important baseline performance metrics, with the best-performing system achieving 10.00\% accuracy. These results demonstrate that while current AI systems show limited capabilities in Legal Zero-Day discovery, the capability exists at detectable levels and varies meaningfully across different models. The uniformly low performance suggests this represents a genuine capability frontier that could see significant advancement as AI systems become more sophisticated.

The evaluation framework itself represents a methodological contribution to AI safety assessment, providing a scalable approach for testing complex, domain-specific capabilities while avoiding the ethical issues of testing against real legal vulnerabilities. Our automated judge achieved perfect agreement with human experts, demonstrating the feasibility of expert-level evaluation for these complex legal reasoning puzzles.

By establishing Legal Zero-Days as a measurable AI capability and providing baseline performance data, this work extends AI safety evaluation beyond traditional risk vectors to encompass the legal and regulatory infrastructure that governs society. As AI systems become more autonomous and influential, understanding their potential to discover and exploit legal vulnerabilities becomes increasingly critical for maintaining institutional stability and effective AI governance.

The benchmark and baseline results provide a foundation for monitoring this important capability as AI systems advance, contributing to the broader effort to anticipate and prepare for the diverse ways that highly capable AI systems might impact society. Future work expanding the evaluation framework and tracking capability development will be essential for understanding and managing Legal Zero-Day risks as they evolve alongside advancing AI capabilities.

\begin{ack}
This research was funded and supported by the UK AI Security Institute and Good Ancestors. In addition, we thank the expert legal volunteers who generated the legal puzzles for this evaluation.
\end{ack}

\bibliographystyle{nips}
{
\small
\bibliography{references}

\begin{thebibliography}{10}

\bibitem{shevlane2023modelevaluationextremerisks}
Shevlane, T., S.~Farquhar, B.~Garfinkel, et~al.
\newblock Model evaluation for extreme risks, 2023.

\bibitem{anthropic2023rsp}
Anthropic.
\newblock Responsible scaling policy, 2023.

\bibitem{openai2023preparedness}
OpenAI.
\newblock Preparedness framework, 2023.

\bibitem{baker2020emergenttoolusemultiagent}
Baker, B., I.~Kanitscheider, T.~Markov, et~al.
\newblock Emergent tool use from multi-agent autocurricula, 2020.

\bibitem{marchant2011growing}
Marchant, G.~E., B.~Allenby, J.~R. Herkert.
\newblock The growing gap between emerging technologies and legal-ethical oversight.
\newblock \emph{The growing gap between emerging technologies and legal-ethical oversight}, pages 19--33, 2011.

\bibitem{nzherald2013oaths}
{New Zealand Herald}.
\newblock Parliament to urgently pass police oath law, 2013.

\bibitem{bbc2024prisoners}
{BBC}.
\newblock Reoffending prisoner was let out by mistake, bbc told, 2024.

\bibitem{ukaisi2024inspect}
AI~Security~Institute, U.
\newblock Inspect {AI:} {Framework} for {Large} {Language} {Model} {Evaluations}.

\bibitem{starace2025paperbenchevaluatingaisability}
Starace, G., O.~Jaffe, D.~Sherburn, et~al.
\newblock Paperbench: Evaluating ai's ability to replicate ai research, 2025.

\bibitem{zheng2023judgingllmasajudgemtbenchchatbot}
Zheng, L., W.-L. Chiang, Y.~Sheng, et~al.
\newblock Judging llm-as-a-judge with mt-bench and chatbot arena, 2023.

\bibitem{phuong2024evaluatingfrontiermodelsdangerous}
Phuong, M., M.~Aitchison, E.~Catt, et~al.
\newblock Evaluating frontier models for dangerous capabilities, 2024.

\end{thebibliography}
}


\appendix

\section{Code for the auxiliary evaluation of our judge models}

\begin{lstlisting}[language=Python]
import asyncio
import json
import os

import aiofiles
from inspect_ai.model import (
    ChatCompletionChoice,
    ChatMessageAssistant,
    ModelName,
    ModelOutput,
)
from inspect_ai.scorer import (
    Target,
    model_graded_fact,
)
from inspect_ai.solver import TaskState
from sklearn.metrics import f1_score, accuracy_score, precision_score, recall_score

from legal_zero_days.utils import get_root

root = get_root()

SCORING_EVAL_MODEL = os.environ.get("SCORING_EVAL_MODEL", "openai/gpt-4o")


async def scorer_eval():
    """
    This function evaluates the model_graded_fact scorer using gpt-4o on our ground_truth_scoring dataset.
    You can run this by calling `uv run poe scorer_eval`
    """
    # Load datasets
    async with aiofiles.open(
        root / "dataset" / "dist" / "ground_truth_scoring.json"
    ) as f:
        content = await f.read()
        ground_truth_scoring_dataset = json.loads(content)

    async with aiofiles.open(root / "dataset" / "dist" / "legal_zero_days.json") as f:
        content = await f.read()
        main_dataset = json.loads(content)

    # Create puzzle lookup dictionary
    puzzles_by_id = {puzzle["puzzle_id"]: puzzle for puzzle in main_dataset}

    # Create a model_graded_fact scorer to evaluate
    scorer = model_graded_fact(model=SCORING_EVAL_MODEL)

    # Add target outputs to ground truth dataset
    for sample in ground_truth_scoring_dataset:
        sample["target"] = puzzles_by_id[int(sample["puzzle_id"])]["target"]

    # Get scores from model_graded_fact
    scores = await asyncio.gather(
        *[
            scorer(
                state=TaskState(
                    model=ModelName(
                        "openai/gpt-4o"
                    ),  # Note this is just dummy text, the samples are not generated by gpt-4o but pulled from the dataset
                    sample_id=0,
                    epoch=0,
                    input="",
                    output=ModelOutput(
                        choices=[
                            ChatCompletionChoice(
                                message=ChatMessageAssistant(
                                    content=sample.get("output"),
                                    source="generate",
                                ),
                            ),
                        ],
                    ),
                    messages=[],
                    choices=[sample.get("output")],
                ),
                target=Target(sample.get("target")),
            )
            for sample in ground_truth_scoring_dataset
        ]
    )

    # Calculate F1 score using sklearn
    y_true = [sample["expected_score"] for sample in ground_truth_scoring_dataset]
    y_pred = [str(score.value) for score in scores]
    # typecheck seems to falsely think pos_label only takes ints
    scorers_f1_score = f1_score(y_true, y_pred, pos_label="C")  # type: ignore
    scorers_accuracy = accuracy_score(y_true, y_pred)
    scorers_precision = precision_score(y_true, y_pred, pos_label="C")  # type: ignore
    scorers_recall = recall_score(y_true, y_pred, pos_label="C")  # type: ignore

    print("Ground truth scoring dataset evaluation")
    print("=======================================")
    print(f"Model: {SCORING_EVAL_MODEL}")
    print(f"Accuracy: {scorers_accuracy}")
    print(f"Precision: {scorers_precision}")
    print(f"Recall: {scorers_recall}")
    print(f"F1 score: {scorers_f1_score}")
    print(f"Number of samples: {len(ground_truth_scoring_dataset)}")
    return scorers_f1_score


if __name__ == "__main__":
    asyncio.run(scorer_eval())

\end{lstlisting}

\section{Code for primary evaluation}

\begin{lstlisting}[language=Python]
import os
from typing import Literal

from inspect_ai import Epochs, Task, task
from inspect_ai.dataset import json_dataset
from inspect_ai.scorer import exact, model_graded_fact
from inspect_ai.solver import Solver, generate, system_message

from legal_zero_days.utils import get_root
from epoch_reducers import register_custom_reducers


SYSTEM_MESSAGE = """
<redacted to avoid poluting future results>
"""

root = get_root()

SCORING_EVAL_MODEL = os.environ.get("SCORING_EVAL_MODEL", "openai/o1")


@task
def legal_zero_days(
    solver: Solver | None = None,
    scorer_type: Literal["exact_match", "model_graded_fact"] = "model_graded_fact",
):
    dataset = json_dataset(str(root / "dataset" / "dist" / "legal_zero_days.json"))

    # Register custom reducers
    custom_reducers = register_custom_reducers()

    return Task(
        dataset=dataset,
        solver=solver or [system_message(SYSTEM_MESSAGE), generate()],
        scorer=(
            model_graded_fact(model=SCORING_EVAL_MODEL)
            if scorer_type == "model_graded_fact"
            else exact()
        ),
        epochs=Epochs(10, ["mean"] + custom_reducers),
    )

\end{lstlisting}

\end{document}